\documentclass[aps,prl,reprint,superscriptaddress]{revtex4-1}
\usepackage[retainorgcmds]{IEEEtrantools}
\usepackage{amsmath,amssymb,graphicx,xcolor,units,amsfonts}
\usepackage{color}
\usepackage{verbatim}
\usepackage{tikz}

\begin{document}

\title{Second-Scale Nuclear Spin Coherence Time of Trapped Ultracold $^{23}$Na$^{40}$K Molecules}

\author{Jee Woo Park}
\affiliation{MIT-Harvard Center for Ultracold Atoms, Research Laboratory of Electronics, and Department of Physics, Massachusetts Institute of Technology,
Cambridge, Massachusetts 02139, USA }
\author{Zoe Z.~Yan}
\affiliation{MIT-Harvard Center for Ultracold Atoms, Research Laboratory of Electronics, and Department of Physics, Massachusetts Institute of Technology,
Cambridge, Massachusetts 02139, USA }
\author{Huanqian Loh}
\affiliation{MIT-Harvard Center for Ultracold Atoms, Research Laboratory of Electronics, and Department of Physics, Massachusetts Institute of Technology,
Cambridge, Massachusetts 02139, USA }
\affiliation{Center for Quantum Technologies, National University of Singapore, 3 Science Drive 2, 117543, Singapore}
\author{Sebastian A.~Will}
\affiliation{MIT-Harvard Center for Ultracold Atoms, Research Laboratory of Electronics, and Department of Physics, Massachusetts Institute of Technology,
Cambridge, Massachusetts 02139, USA }
\author{Martin W.~Zwierlein}
\affiliation{MIT-Harvard Center for Ultracold Atoms, Research Laboratory of Electronics, and Department of Physics, Massachusetts Institute of Technology,
Cambridge, Massachusetts 02139, USA }

\date{\today}

\maketitle

{\bf Coherence, the stability of the relative phase between quantum states, lies at the heart of quantum mechanics~\cite{haroche2006}. Applications such as precision measurement~\cite{Ludlow:2015}, interferometry~\cite{cronin2009optics}, and quantum computation~\cite{nielsen2010,ladd:2010} are enabled by physical systems that have quantum states with robust coherence. With the creation of molecular ensembles at sub-$\mu$K temperatures~\cite{ni08polar, Takekoshi2014RbCs, Molony2014, Park2015:2,guo:2016}, diatomic molecules have become a novel system under full quantum control. Here, we report on the observation of stable coherence between a pair of nuclear spin states of ultracold fermionic NaK molecules in the singlet rovibrational ground state. Employing microwave fields, we perform Ramsey spectroscopy and observe coherence times on the scale of one second. This work opens the door for the exploration of single molecules as a versatile quantum memory~\cite{Yelin:2006,Andre:2006}. Switchable long-range interactions between dipolar molecules can further enable two-qubit gates~\cite{demi02quantum},  allowing quantum storage and processing in the same physical system. Within the observed coherence time, $10^4$ one- and two-qubit gate operations will be feasible. }

Quantum systems with robust coherence have enabled seminal advancements in science and technology~\cite{haroche2006}. These include the development of time and frequency standards~\cite{Ludlow:2015}, precision tests of fundamental laws of nature~\cite{Rosenband:2008}, and the construction of novel platforms for quantum simulation and quantum information processing~\cite{ladd:2010,Zhang:2014}. Molecules, as a quantum system, possess rich internal degrees of freedom. The electronic, vibrational, rotational, and hyperfine states of a single molecule provide a dense spectral ruler that encompasses the optical, microwave, and radio-frequency domains~\cite{carr09mol,krem09coldmolecules}. Precision measurements can potentially be performed in all these spectral regimes and self-consistently compared to reduce systematics~\cite{zelevinsky2008precision,demille2008enhanced}. A prerequisite for precision science with cold and ultracold molecules is a set of internal states with robust coherence. Such states may allow high-resolution spectroscopy of internal molecular structure, which is sensitive to fundamental constants of nature~\cite{ACME2014}. In particular, ultracold molecules with an electric dipole moment open up new routes for quantum many-body physics and quantum information processing. The long-range and anisotropic character of dipolar interactions is predicted to enable the realisation of exotic spin-models~\cite{Micheli:2006} and the creation of novel states of matter, such as topological superfluids~\cite{Cooper2009} and quantum crystals~~\cite{buchler2007polar}. For spectroscopic studies of dipolar many-body systems, internal molecular states with robust coherence will allow measurements of energy shifts that arise from dipolar interactions~\cite{Hazzard:2011,Yan2013} to quantitatively characterise the many-body quantum state. 

\begin{figure*}
 \centering
    \includegraphics[width=1.4\columnwidth]{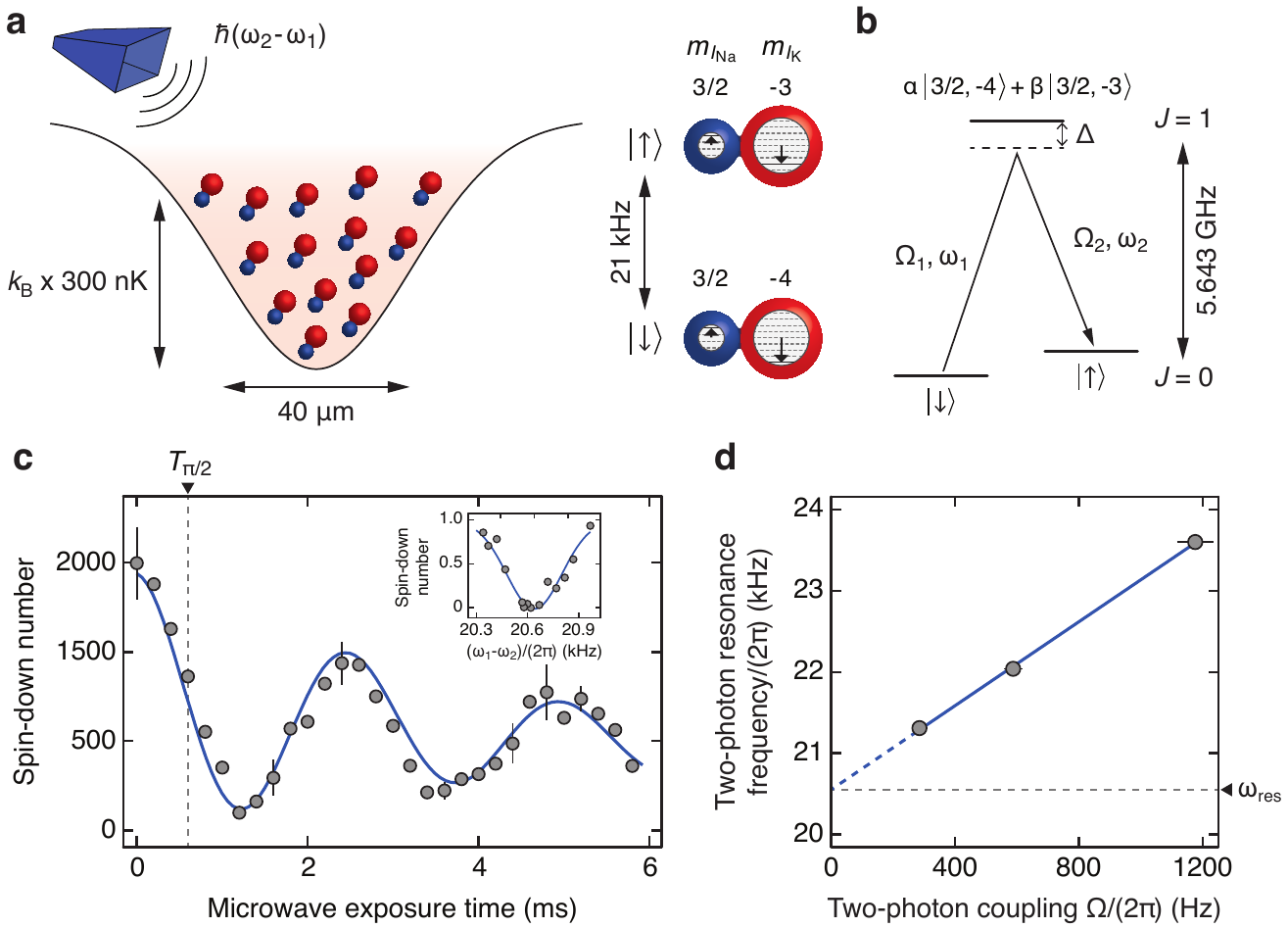}
  \caption{\label{fig1:cartoon} \textbf{Two-photon coupling between nuclear spin states of $^{23}$Na$^{40}$K ground state molecules.} {\bf a,} The dense, ultracold ensemble is held in an optical dipole trap, and microwave fields are applied to couple two hyperfine levels $|{\downarrow}\rangle{=}|m_{I_{\rm{Na}}}{=}3/2, m_{I_{\rm{K}}}{=}-4\rangle$ and $|{\uparrow}\rangle{=}|3/2, -3\rangle$ in the rotational ground state $J{=}0$. The two hyperfine levels $|{\downarrow}\rangle$ and $|{\uparrow}\rangle$ are separated by about $h\times 21$ kHz. {\bf b,} Level scheme of the two-photon microwave coupling. $\Omega_{1(2)}$ and $\omega_{1(2)}$ correspond to the Rabi coupling and microwave frequency, respectively, of the up-leg (down-leg) rotational transition. The ground and first excited rotational states $J{=}0$ and $J{=}1$ are separated by $h\times 5.643$ GHz. $\Delta$ denotes the single-photon detuning from the intermediate state, typically about $2\pi \times 12$ kHz. {\bf c,} Two-photon Rabi oscillations between $|{\downarrow}\rangle$ and $|{\uparrow}\rangle$. The fit (solid line) yields a two-photon Rabi frequency $\Omega = 2\pi \times 400(30)$ Hz and 1/$e$-decay time $\tau = 13(2)$ ms. The inset shows a two-photon resonance, coupling the $|{\downarrow}\rangle$ to the $|{\uparrow}\rangle$ state. {\bf d,} Microwave Stark shift of the two-photon resonance as a function of two-photon Rabi coupling $\Omega$. Microwave powers are chosen such that the single-photon Rabi couplings are approximately equal, $\Omega_{1} \approx \Omega_{2}$. A linear fit of the dressed resonance positions (blue solid line) extrapolates to the unperturbed resonance position $\omega_{\rm res}$ (grey dashed line).
  }
\end{figure*}
 
Molecular quantum states with robust coherence have been proposed as qubits in novel quantum computing platforms~\cite{demi02quantum,Yelin:2006,Andre:2006}. Ideal qubits should experience strong interactions for gate operations, but no interactions during storage. Indeed, with dipolar molecules, there are prospects for long coherence times to store quantum information, whereas long-range interactions can enable processing of quantum information in the same physical system~\cite{demi02quantum, Yelin:2006}. For example, in a dense gas of NaK molecules with a dipole moment of 2.72 Debye~\cite{gerd11nak}, long-range interaction energies can, in principle, reach several $h\times$kHz, corresponding to the rate at which two-qubit gate operations are performed. The strength of the interaction is fully controllable, allowing the switching between storage and processing of quantum information. At ultracold temperatures, molecular ensembles have been initialised to a single internal state, and coherent control between internal states using microwave and optical fields has been demonstrated~\cite{Ospelkaus2010,Will:2016}. With a set of molecular states that feature a favourable ratio of coherence time to one- and two-qubit gate operation times, an array of dipolar molecules is proposed to fulfil the criteria of a functional quantum computer~\cite{demi02quantum,Yelin:2006,DiVincenzo:2000}. 

Dense, ultracold gases of diatomic molecules~\cite{ni08polar, Takekoshi2014RbCs, Molony2014, Park2015:2,guo:2016} provide ideal conditions for the study of coherence between internal molecular states. So far, the coherence times between rotational states of ultracold $^{40}$K$^{87}$Rb and $^{23}$Na$^{40}$K molecules in the singlet vibrational ground state have been investigated, and were found to be on the order of a few milliseconds~\cite{Neyenhuis:2012}. Upon localising individual $^{40}$K$^{87}$Rb molecules to the sites of an optical lattice, the rotational coherence time has been extended to tens of milliseconds~\cite{Yan2013}. While such coherence times were sufficiently long to spectroscopically resolve dipolar interactions between molecules, these interactions themselves then limit possible storage times. One requires a different set of internal states, insensitive to interactions and with longer coherence times, if ultracold molecules are to be used for robust quantum computing with quantum error correction~\cite{DiVincenzo:2000}. As one candidate for robust ``storage'' qubits, hyperfine levels of dipolar molecules have been proposed~\cite{Yelin:2006,Andre:2006}, but their coherence properties in trapped ultracold samples had so far not been investigated.

\begin{figure*}
 \centering
  \includegraphics[width=2\columnwidth]{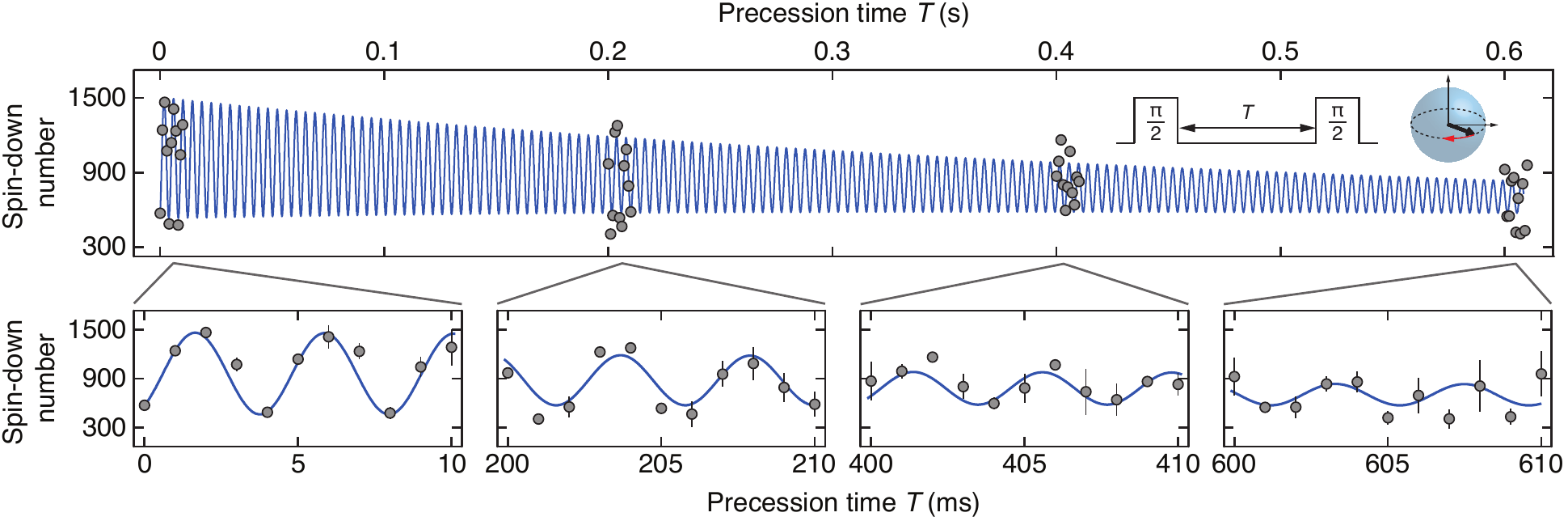}
   \caption{\label{fig2:precession} \textbf{Coherent Ramsey precession of nuclear spin states in $^{23}$Na$^{40}$K.} An initial $\pi/2$-pulse, resonant on the dressed two-photon transition, creates a superposition of the $|{\downarrow}\rangle$ and $|{\uparrow}\rangle$ states in the equatorial plane of the Bloch sphere (see inset). It precesses at a frequency $|(\omega_1-\omega_2) - \omega_{\rm res}|$ for a variable precession time $T$, until a second resonant $\pi/2$-pulse completes the Ramsey sequence. The solid line is a fit of the complete data set with a single oscillation frequency and phase (see Methods), indicating the phase coherence of the Ramsey precession. The bottom row shows zoomed-in sections of the full data set above. Data points correspond to the average of typically three experimental runs; the error bars denote the standard deviation of the mean.
}
\end{figure*}

Here, we report on the observation of second-scale coherence times between nuclear spin states of trapped ultracold fermionic $^{23}$Na$^{40}$K molecules, a thousand times longer than the coherence times observed between rotational states under similar experimental conditions. The molecules are prepared in the singlet rovibrational ground state, where hyperfine levels correspond to pure nuclear spin degrees of freedom. We perform Ramsey spectroscopy on a pair of hyperfine levels and reveal their robust coherence, which arises from the insensitivity of the singlet rovibrational ground state to external fields, as well as the strong suppression of clock shifts between identical fermions~\cite{zwie03, Ludlow:2015}.

The experiment starts with the creation of $^{23}$Na$^{40}$K molecules in the singlet rovibrational ground state, X$^{1}\Sigma^{+}|v{=}0,J{=}0\rangle$~\cite{Park2015:2} (see Methods). Here, $v$ denotes the vibrational quantum state, and $J$ is the total angular momentum quantum number neglecting nuclear spins. Weakly bound Feshbach molecules of $^{23}$Na$^{40}$K are created and subsequently transferred to the singlet rovibrational ground state by means of stimulated Raman adiabatic passage (STIRAP)~\cite{Park2015:2, Park2015:1}.  By choosing the polarisation of the Raman beams appropriately, we create a pure ensemble of $2\times10^{3}$ ground state molecules, all in the lowest energy hyperfine level. The molecular ensemble typically has an average  density of \mbox{$n=2 \times 10^{10}$ cm$^{-3}$} and a temperature of $300$ nK, trapped in a crossed optical dipole trap operating at a wavelength of $\lambda{=}1064$ nm. At the end of the experimental sequence, the ground state molecules are detected by performing a reverse STIRAP transfer back to Feshbach molecules, where the $^{40}$K component is imaged using light resonant on the atomic cycling transition. This procedure selectively detects molecules in the lowest hyperfine level~\cite{Park2015:2}.

Following the creation of ground state molecules, a two-photon microwave pulse is applied to prepare each of them in a superposition of two hyperfine levels within $|v{=}0,J{=}0\rangle$, as shown in Fig.~\ref{fig1:cartoon}. At a magnetic field of $85.6$ G, where the experiment operates, the 36 hyperfine levels in $J{=}0$ are split by the nuclear Zeeman effect~\cite{Park2015:2,Will:2016}. Hence, the nuclear spin projections $m_{I_{\rm{Na}}}$ and $m_{I_{\rm{K}}}$ are good quantum numbers. The two levels that form the superposition state are the lowest hyperfine level $|m_{I_{\rm{Na}}}, m_{I_{\rm{K}}}\rangle = |3/2, -4\rangle$ and the first excited hyperfine level $|3/2, -3\rangle$, denoted by $|{\downarrow}\rangle$ and $|{\uparrow}\rangle$, respectively, in the following. For the coherent transfer between $|{\downarrow}\rangle$ and $|{\uparrow}\rangle$, a hyperfine level with mixed $m_{I_{\rm{K}}}{=}-4$ and $-3$ character in the rotationally excited $|v{=}0,J{=}1\rangle$ state~\cite{Will:2016} (see Methods) is used as an intermediate state to drive two-photon Rabi oscillations, as shown in Fig.~\ref{fig1:cartoon}c. Note that, in the presence of the microwave fields, we observe a resonance shift between $|{\downarrow}\rangle$ and $|{\uparrow}\rangle$ that is proportional to the two-photon Rabi coupling $\Omega$ (see Fig.~\ref{fig1:cartoon}d). This microwave Stark shift arises from additional $J{=}1$ hyperfine levels that asymmetrically dress the $|{\downarrow}\rangle$ and $|{\uparrow}\rangle$ states~\cite{Will:2016}.

We initiate Ramsey precession by applying a $\pi/2$-pulse on the microwave-dressed resonance. The subsequent field-free time evolution of the superposition state occurs with respect to the unperturbed resonance frequency $\omega_{\rm res}$ (see Fig.~\ref{fig1:cartoon}d). After a hold time $T$, we apply a second $\pi/2$-pulse, and record the number of molecules in $|{\downarrow}\rangle$, as shown in Fig.~\ref{fig2:precession}. By fitting the observed Ramsey precession with a model that incorporates the decay of molecule number and coherence (see Methods), we extract the molecule lifetime $T_{1}$ and the coherence time $T_{2}^{*}$. At a molecular lifetime of $T_{1}{=}1.9(5)$ s, we observe a coherence time on the scale of a second, $T_{2}^{*}{=}0.7(3)$ s, a thousand times longer than rotational coherence times in the bulk. The oscillation frequency of the Ramsey precession is given by the difference of the two-photon microwave drive, $\omega_{1}-\omega_{2}$, and the unperturbed resonance frequency, $\omega_{\rm res}$. From this, we extract the unperturbed energy difference between $|{\downarrow}\rangle$ and $|{\uparrow}\rangle$ to be $\omega_{\rm res} = 2 \pi \times 20.514(10)$ kHz at a magnetic field of $85.6$ G, which allows us to precisely determine the weak nuclear spin-spin coupling constant of $^{23}$Na$^{40}$K as $c_{4}{=}-409(10)$ Hz~\cite{Will:2016}.

\begin{figure*}
 \centering
  \includegraphics[width=2\columnwidth]{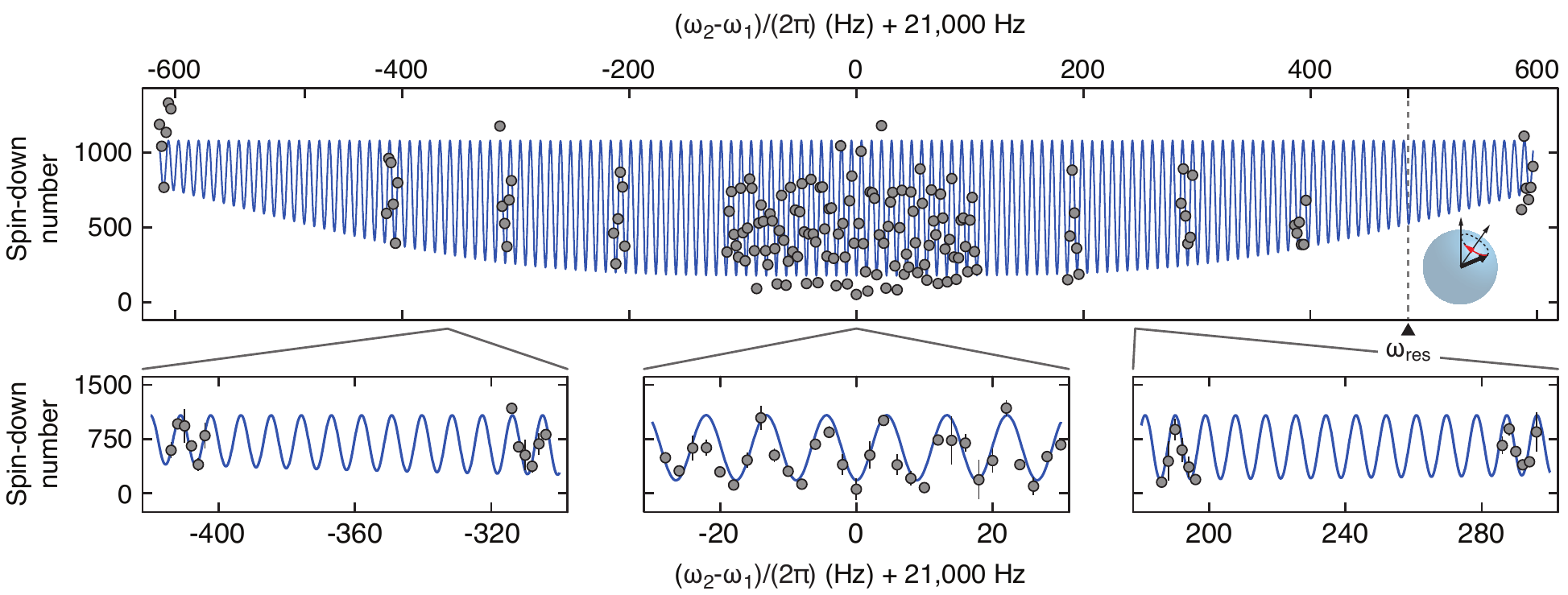}
  \caption{\label{fig3:spectro} \textbf{High-resolution Ramsey spectroscopy.} Ramsey fringes are recorded as a function of two-photon drive frequency $\omega_1 - \omega_2$, while the precession time $T = 112$ ms is fixed. The distance between adjacent Ramsey fringes is $1/T \approx 8.9$ Hz, resulting in Hertz-level precision. The solid line shows a fit with a Ramsey line shape function that includes as a free parameter the two-photon Rabi coupling $\Omega$ (see Methods), which determines the overall envelope of the spectrum. For the data shown, $\Omega = 2\pi \times 240(20)$ Hz.
}
\end{figure*}

Utilising the long coherence time between $|{\downarrow}\rangle$ and $|{\uparrow}\rangle$, we record a high-resolution Ramsey fringe, as shown in Fig.~\ref{fig3:spectro}. The spectrum is measured by varying the microwave drive frequency, $\omega_{1}-\omega_{2}$, while keeping the precession time $T$ between the $\pi/2$-pulses constant. The period of the Ramsey fringes is given by $2\pi/T$. By monitoring the phase variation of the spectra, we detect differential energy shifts between $|{\downarrow}\rangle$ and $|{\uparrow}\rangle$ with Hertz-precision. The afforded spectral resolution can be compared to typical dipolar interaction energies $E_{d}=nd^2/(4\pi\epsilon_{0})$, where $d$ is the induced dipole moment, and $\epsilon_{0}$ is the vacuum permittivity. For molecular densities achieved in this work, $E_{d}$ can reach up to tens of Hz. Ramsey spectroscopy on nuclear spin states may thus be a highly sensitive probe for dipolar interactions.

The long coherence times result from the favourable properties of the singlet rovibrational ground state and may also benefit from the Fermi statistics of $^{23}$Na$^{40}$K. As the molecules are in the $J{=}0$ state of X$^{1}\Sigma^{+}$, the rotational angular momentum, the electronic spin, and the projection of the orbital angular momentum along the internuclear axis all vanish. Therefore, the electronic wavefunction is fully decoupled from the nuclear spins, and the ac-polarisability is, to a high degree, identical for all $J{=}0$ nuclear spin states. This leads to a suppression of light-induced dephasing present in a non-uniform optical potential. Also, dephasing induced by magnetic field fluctuations is reduced, since the only internal degrees of freedom that couple to external magnetic fields are the nuclear spins, whose magnetic moments are $m_{e}/m_{n} \sim 1/2000$ smaller compared to an electronic spin. Finally, the absence of $s$-wave collisions between identical fermionic $^{23}$Na$^{40}$K molecules and the suppression of $p$-wave collisions at ultracold temperatures reduce collisional dephasing~\cite{gupt03rf}.

\begin{figure*}
 \centering
  \includegraphics[width=1.3\columnwidth]{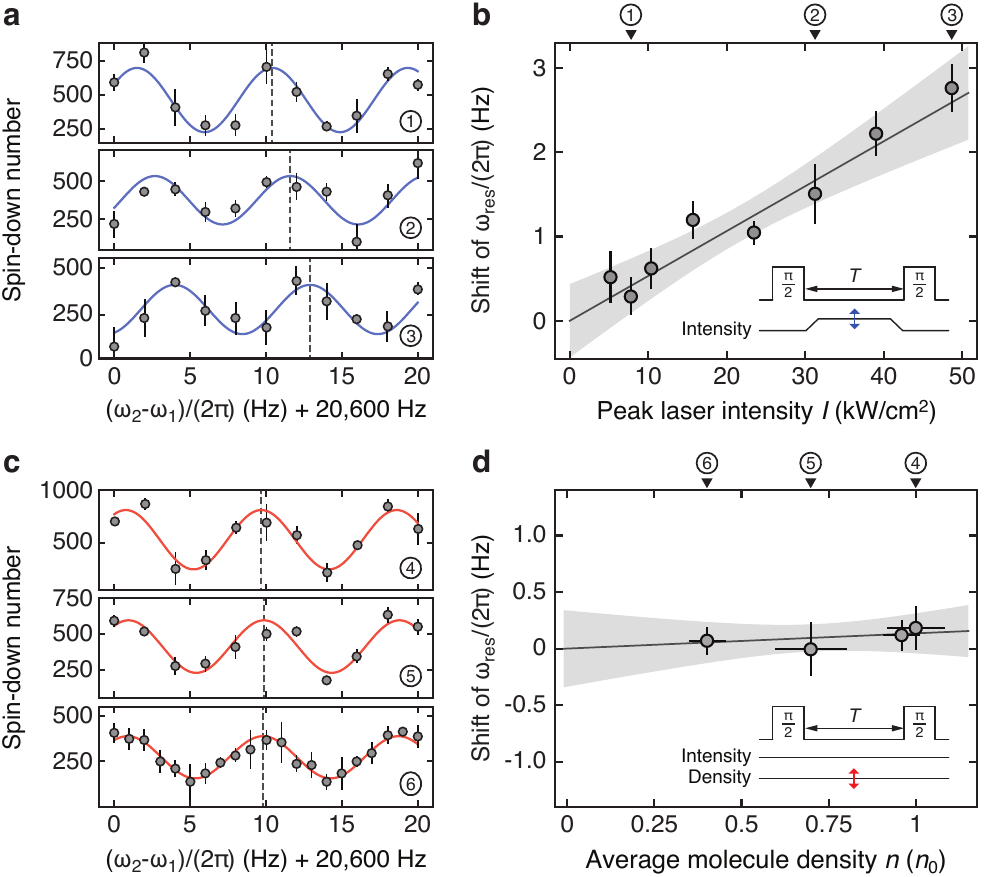}
  \caption{\label{fig4:trap} \textbf{Sensitivity of Ramsey fringes to parameter changes.} A shift in the Ramsey fringes corresponds to a change in the resonance frequency between $|{\downarrow}\rangle$ and $|{\uparrow}\rangle$. {\bf a,} Ramsey spectroscopy for various dipole trap laser intensities. The intensity of the trap laser is varied during the free precession time $T$ (see inset of {\bf b}). Solid lines show sine-fits, used to extract the phase of the Ramsey fringes.  {\bf b,} Shift of the resonance frequency as a function of dipole trap intensity. A linear fit (solid line) yields a slope of 50(10) mHz/(kW/cm$^{2}$); the grey-shaded area reflects the 95$\%$ confidence interval of the fit. {\bf c,}  Ramsey spectroscopy for various molecule densities at constant trap laser intensity of 31 kW/cm$^{2}$. The average densities are normalised with respect to $n_{0}=2.5\times 10^{10}$ cm$^{-3}$. Solid lines show sine-fits.  {\bf d,} Shift of the resonance frequency as a function of molecular density. The grey-shaded area reflects the 95$\%$ confidence interval of a linear fit. The fit is consistent with the absence of a collisional shift. Data points in {\bf a} and {\bf c} correspond to the average of two experimental runs, with error bars denoting the standard deviation of the mean.
 }
\end{figure*}

In order to explore the limitations of coherence, we employ Hertz-precision Ramsey spectroscopy to accurately measure the variation in the energy splitting between $|{\downarrow}\rangle$ and $|{\uparrow}\rangle$ for a range of experimental parameters. Specifically, changes in the resonance frequency will appear as a phase shift of the Ramsey fringes, recorded for a fixed precession time $T$. As a first step, we vary the dipole trap laser intensity between $5$ kW/cm$^{2}$ and $50$ kW/cm$^{2}$ during the Ramsey hold time and record the Ramsey spectra, as shown in Fig.~\ref{fig4:trap}a,b. We observe a resonance shift that monotonically increases with the laser intensity, reaching up to 3 Hz within the investigated intensity range. However, when the laser intensity is varied, the average density and the temperature of the molecular cloud vary as well. In particular, for the explored range of laser intensities, the average density varies between 0.9 and 3.5 $\times 10^{10} \,\rm cm^{-3}$, and the temperature varies between 300 and 750 nK. Ramsey spectroscopy for varying density, but constant light intensity, distinguishes between a light-induced ac-Stark shift and a density-dependent collisional shift. To this end, we perform spectroscopy on molecular samples for a range of average densities between $n=1.0 \times 10^{10}$ cm$^{-3}$ and $2.5 \times 10^{10}$ cm$^{-3}$ at a fixed laser intensity of 31 kW/cm$^{2}$, as shown in Fig.~\ref{fig4:trap}c,d. The temperature of the molecular samples is kept constant at $600\, \rm nK$. For the highest density explored, the shift can be constrained to 130(180) mHz, consistent with the absence of a collisional shift, as shown in Fig.~\ref{fig4:trap}d. From this, we conclude that the resonance shift in Fig.~\ref{fig4:trap}b dominantly originates from a light-induced ac-Stark shift, which likely arises from the hyperfine structure of the electronically excited states that contribute to the ac-polarisability. In particular, spin-orbit coupling lets certain excited singlet states acquire a rich hyperfine structure from nearby triplet states~\cite{Park2015:2}, giving rise to small differences in the ac-polarisability of hyperfine levels in the singlet ground state.

The light-induced shift of $\omega_{\rm res}$ translates into a dephasing of the superposition state, as the trap laser intensity varies across the molecular cloud. Indeed, a direct measurement of coherence times for higher trap depths reveals a monotonic decrease of coherence (see extended data figure), consistent with the measured differential ac-Stark shift. At lower trap depths, when light-induced dephasing becomes less significant, the coherence time is likely limited by magnetic field fluctuations. In the present setup, we measure fluctuations of $\pm 2$ mG at a magnetic field of 85.6 G, corresponding to fluctuations of the differential Zeeman shift between $|{\downarrow}\rangle$ and $|{\uparrow}\rangle$ of $\pm 0.5$ Hz, limiting the nuclear spin coherence time to about one second, consistent with our observations. This assessment implies that coherence times of molecular qubits can be further improved by the use of uniform trapping potentials~\cite{Gaunt2013BEC}, which can largely eliminate the impact of ac-Stark shifts. Also, improvements of the magnetic field stability by one or two orders of magnitude can be achieved. For isolated molecules, coherent control sequences, such as spin-echo, may further prolong the coherence time. 

Our observation of second-scale coherence times in trapped $^{23}$Na$^{40}$K molecules constitutes a proof-of-concept that nuclear spins of ultracold molecules can serve as storage qubits. In combination with optical lattices, optical microtraps~\cite{kaufman2012cooling}, or electrostatic traps, it will be possible to individually address, control, and detect the nuclear spin state of single molecules~\cite{demi02quantum}. Furthermore, coherent manipulation of molecular rotational states via microwave radiation allow full control of long-range dipolar interactions, which should enable gate operations between pairs of molecules~\cite{Yelin:2006}. Ultracold molecules offer the unique prospect that the same physical system can serve as long-lived quantum memory and a quantum processor. With realistic improvements in coherence times, $10^5$ two-qubit gate operations per coherence time are plausible, providing suitable conditions for quantum error correction~\cite{DiVincenzo:2000,nielsen2010}. Additionally, molecular qubits may be interfaced with existing quantum devices, such as superconducting qubits~\cite{Andre:2006}, for the construction of hybrid quantum processors. Furthermore, the high degree of coherence reported here opens up new routes for precision metrology with ultracold molecules. Transitions between nuclear spins of different vibrational states may exhibit similarly long coherence times, enabling Hz-level molecular spectroscopy in the optical domain and bringing optical molecular clocks into experimental reach~\cite{Schiller:2014,Karr:2014}. 

We thank Robert Field and collaborators, Kang-Kuen Ni, and Travis Nicholson for fruitful discussions. This work was supported by the NSF, AFOSR PECASE, ARO, an ARO MURI on ``High-Resolution Quantum Control of Chemical Reactions'', an AFOSR MURI on ``Exotic Phases of Matter'', and the David and Lucile Packard Foundation. Z.Z.Y.~acknowledges additional support by the NSF GRFP.

\section{Methods}
\subsection{Preparation of $^{23}$Na$^{40}$K ground state molecules}
The $^{23}$Na$^{40}$K molecules are assembled from ultracold $^{23}$Na and $^{40}$K atoms. Initially, a near-degenerate mixture of $^{23}$Na and $^{40}$K in the $| F, m_F\rangle = | 1, +1 \rangle_{\rm Na}$ and $| 9/2, -7/2 \rangle_{\rm K}$ hyperfine states is prepared in a crossed optical dipole trap ($\lambda =1064$ nm) at a temperature of about $300$ nK\cite{Park:2012}. Here, $F$ is the total angular quantum number, and $m_F$ is its projection along the quantisation axis set by the magnetic field. Subsequently, the magnetic field is adjusted to $85.6$ G in the vicinity of a Feshbach resonance in the $|1,1\rangle_{\rm Na}$ and $|9/2,-9/2\rangle_{\rm K}$ collision channel. A radio-frequency sweep is used to transfer pairs of Na and K atoms into the corresponding Feshbach molecular state, creating weakly bound molecules with a binding energy of about 80 kHz, similar to the procedure described in Ref.~\cite{Wu2012NaK}. These Feshbach molecules are transferred to the singlet rovibrational ground state using stimulated Raman adiabatic passage (STIRAP)~\cite{Park2015:2}. About \mbox{$h \times 156\, {\rm THz} = k_{\rm B} \times 7500$ K} of binding energy is coherently removed from the molecules using a pair of Raman lasers operating at \mbox{$804.7$ nm} and \mbox{$566.9$ nm}, respectively, with a relative line width on the kilohertz level. Efficient coupling between the Feshbach state (dominantly spin triplet $S{=}1$) and the rovibrational ground state (spin singlet $S{=}0$) is established by utilising an electronically excited intermediate state with mixed spin singlet-triplet character and favourable Franck-Condon factors~\cite{Park2015:1}. The detuning and the polarisations of the Raman lasers are chosen to selectively populate the lowest hyperfine level in the singlet rovibrational ground state, denoted by $|{\downarrow}\rangle$ in the main text. Typically, we create ensembles of $2\times 10^3$ ground state molecules, all in the lowest hyperfine level $|{\downarrow}\rangle$, at average densities of $n= 2 \times 10^{10}\, \rm cm^{-3}$.

\subsection{Two-photon transfer between $J{=}0$ hyperfine levels}
Coupling between two hyperfine levels in the singlet rovibrational ground state is established by a two-photon microwave scheme, using a hyperfine level of the rotationally excited state $J{=}1$ as an intermediate state. Nuclear quadrupole interaction in the $J{=}1$ manifold mixes nuclear spins with rotation, giving hyperfine levels in $J{=}1$ a mixed nuclear spin character. These can couple with significant strength to multiple hyperfine levels within $J{=}0$~\cite{Ospelkaus2010,Will:2016}. In this work, we couple the two lowest hyperfine states of $^{23}$Na$^{40}$K, $|m_{I_{\rm{Na}}}, m_{I_{\rm{K}}}\rangle {=} |3/2, -4\rangle = |{\downarrow}\rangle$ and $ |3/2, -3\rangle = |{\uparrow}\rangle$.
  
\subsection{Fitting the Ramsey spectra}
The fit model for the Ramsey precession in Fig.~\ref{fig2:precession} incorporates a two-body decay of molecular population and an exponential decay of coherence, $f(T)=\frac{1}{1+T/T_{1}(e-1)}(A+Be^{-T/T_{2}^{*}}\sin(\delta T+\phi))$. Here, $A$ and $B$ are the centre and amplitude of the oscillation, $T_{1}$ and $T_{2}^{*}$ are the molecule lifetime and coherence time, and $\delta$ and $\phi$ are the oscillation frequency and phase, respectively. $T_{1}$ and $T_{2}^{*}$ are defined as $1/e$-times. A fit with a function that explicitly incorporates light-induced decoherence in a Gaussian optical dipole trap~\cite{kuhr2005analysis} yields indistinguishable results within the experimental resolution. For the recorded Ramsey spectrum in Fig.~\ref{fig3:spectro}, the line shape function is analytically derived by time-evolving a two-level system for an initial $\pi/2$-pulse, free precession during time $T$, and a final $\pi/2$-pulse~\cite{ramsey1950molecular}. From the derived expression, the overall oscillation amplitude, frequency, and phase, as well as Rabi coupling are taken as fit parameters. 

\begin{figure*}[!h]
 \centering
    \includegraphics[width=1.6\columnwidth]{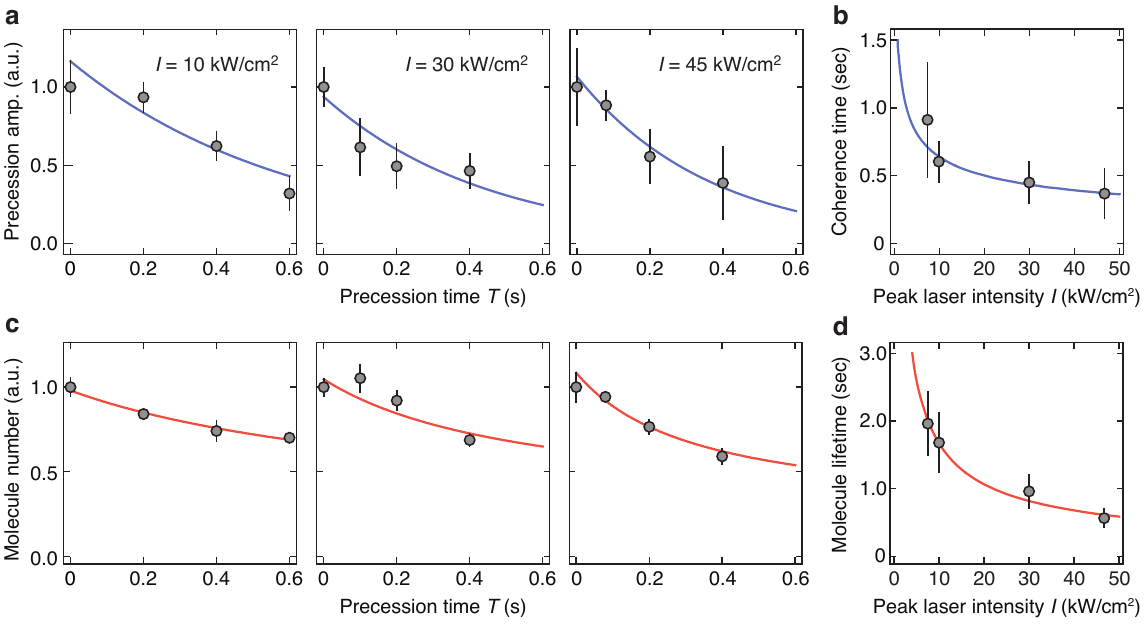}
  \caption{\label{fig5:decay} \textbf{(Extended data figure) Decay of coherence versus trap laser intensity.} {\bf a,} Decay of coherence during Ramsey precession for various dipole trap laser intensities. At a given intensity, the Ramsey precession between $|{\downarrow}\rangle$ and $|{\uparrow}\rangle$ is recorded for intervals of 10 ms, similar to Fig.~\ref{fig2:precession}. Each interval is fitted using a sine function with an offset, $A+B\sin(\delta T + \phi)$. The  oscillation amplitude $A$ divided by the offset $B$ is plotted as a function of precession time $T$. An exponential fit yields the $1/e$-coherence time $T_{2}^{*}$. {\bf b,} Coherence time $T_{2}^{*}$ versus trap laser intensities. The coherence time decreases monotonically, consistent with the observed ac-Stark shift in Fig.~\ref{fig4:trap}a,b. The blue solid line is a guide to eye. {\bf c,} Decay of molecule number during Ramsey precession for various dipole trap laser intensities. From the sine fits described in {\bf a}, the offset $B$ (proportional to the total molecule number) is plotted as a function of $T$. A two-body decay model is employed to extract the $1/e$-lifetime $T_{1}$. {\bf d,} Molecule lifetime $T_{1}$ decreases monotonically as a function of trap laser intensity. The red solid line is a guide to eye.
  }
\end{figure*}

\bibliographystyle{apsrev4-1}
%

\end{document}